\documentclass[twocolumn,prb,notitlepage,floats,superscriptaddress,amsmath,amssymb]{revtex4-1}

\usepackage{bm}
\usepackage[utf8]{inputenc}
\usepackage[T1]{fontenc}
\usepackage{graphicx}
\usepackage{upgreek}
\usepackage{color}
\usepackage{ulem}

\def \FUW{Institute of Experimental Physics, Faculty of Physics, University of Warsaw, ul. Pasteura 5, 02-093 Warsaw, Poland}
\def \LNCMI{Laboratoire National des Champs Magn\'etiques Intenses, CNRS-UGA-UPS-INSA-EMFL, 25, avenue des Martyrs, 38042 Grenoble, France} 
\def \Brno{Central European Institute of Technology, Brno University of Technology,  Purky\v{n}ova 656/123, 612 00 Brno, Czech Republic}

\begin{document}

\title{The effect of metallic substrates on the optical properties of monolayer MoSe$_{2}$}

\author{M.~Grzeszczyk}
\email{Magdalena.Grzeszczyk@fuw.edu.pl}\affiliation{\FUW}
\author{M.~R.~Molas}\affiliation{\FUW}\affiliation{\LNCMI}
\author{K.~Nogajewski}\affiliation{\FUW}\affiliation{\LNCMI}
\author{M.~Barto\v{s}}\affiliation{\LNCMI}\affiliation{\Brno}
\author{A.~Bogucki}\affiliation{\FUW}
\author{C.~Faugeras}\affiliation{\LNCMI}
\author{P.~Kossacki}\affiliation{\FUW}
\author{A.~Babi\'nski}\affiliation{\FUW}
\author{M.~Potemski}\affiliation{\FUW}\affiliation{\LNCMI}

\date{\today}

\begin{abstract}

Atomically thin materials, like semiconducting transition metal dichalcogenides (S-TMDs), are highly sensitive to the environment. This opens up an opportunity to externally control their properties by changing their surroundings. We investigate the effect of several metallic substrates on the optical properties of MoSe$_2$ monolayer (ML) deposited on top of them with photoluminescence and reflectance contrast techniques. The optical spectra of MoSe$_{2}$ MLs deposited on Pt, Au, Mo and Zr have distinctive metal-related lineshapes. In particular, a substantial variation in the intensity ratio and the energy separation between a negative trion and a neutral exciton is observed. It is shown that using metals as substrates affects the doping of S-TMD MLs. The explanation of the effect involves the Schottky barrier formation at the interface between the MoSe$_{2}$ ML and the metallic substrates. The alignment of energy levels at the metal/semiconductor junction allows for the transfer of charge carriers between them. We argue that a proper selection of metallic substrates can be a way to inject appropriate types of carriers into the respective bands of S-TMDs. 
\end{abstract}

\maketitle

\section{Introduction \label{sec:Intro}}


\begin{figure}[t]
\includegraphics[width=\linewidth]{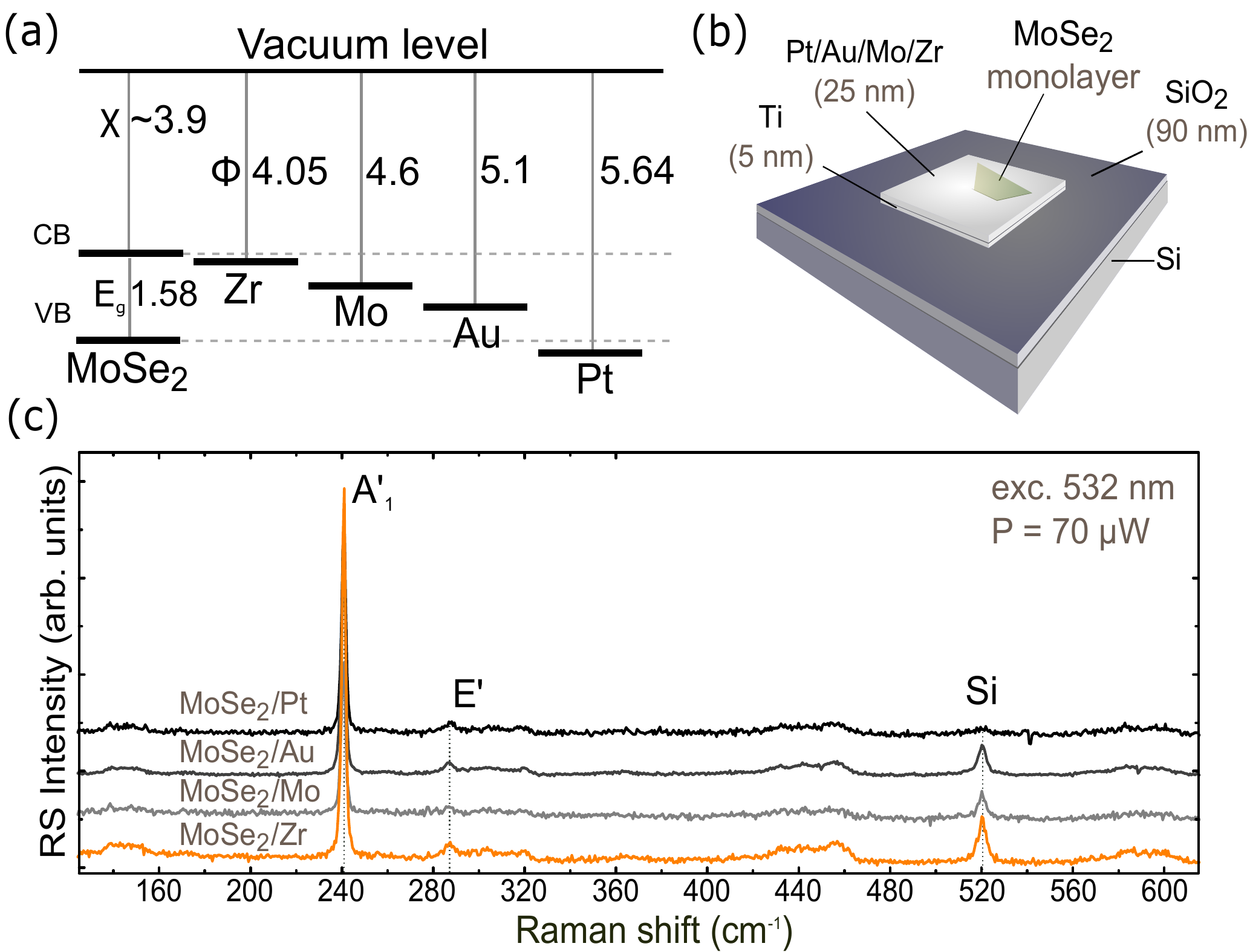}
\caption{(a) Scheme of energy levels diagram of considered MoSe$_{2}$/metal heterostructures. $\chi$ and $\Phi$ denote the electron affinity and metal work function, respectively. CB and VB mark the conduction and valence bands, E$_{g}$ is the energy band gap of monolayer MoSe$_{2}$. All values are given in electronvolts (eV). (b) Schematic illustration of samples under study. (c) Room-temperature Raman scattering spectra of monolayer MoSe$_{2}$ on different metallic substrates.}
\label{fig:probka}
\end{figure}

Out-of-plane quantum confinement in monolayers (MLs) of semiconducting transition metal dichalcogenides (\mbox{S-TMDs}), as well as their large in-plane effective masses of electrons and holes contribute to strong Coulomb interactions between charge carriers, which is reflected in large exciton binding energies. \cite{mak2010atomically, ramasubramaniam2012large} Due to the nature of these materials, their electronic and optical properties are highly sensitive to their surroundings. This effect can be used as a non-invasive way to influence the screening of electron-hole Coulomb interaction in S-TMDs MLs \cite{coulomb,screening,gupta2017direct,steinke2017non,rosner2016two}. On the other hand, the electronic properties of atomically thin layers can be locally altered by metals, which are contacted with the samples \cite{kang2012computational, cakir2014dependence, duan2015two}. In consequence, using metals as substrates may affect doping levels in S-TMD MLs due to the alignment of energy bands at the metal/semiconductor junctions. A selection of suitable substrates can be a way to inject appropriate types of carriers into the respective bands of S-TMDs. The efforts to better understand the role of interfaces and doping processes may be of great value for future applications of thin S-TMD layers in a variety of modern electronic devices (field-effect transistors \cite{FET}, sensors \cite{sensor}, spintronic\cite{spintronic} and valleytronic circuits\cite{valleytronic} etc.) since all of them incorporate metallic contacts.

\begin{figure*}[t]
\includegraphics[width=\linewidth]{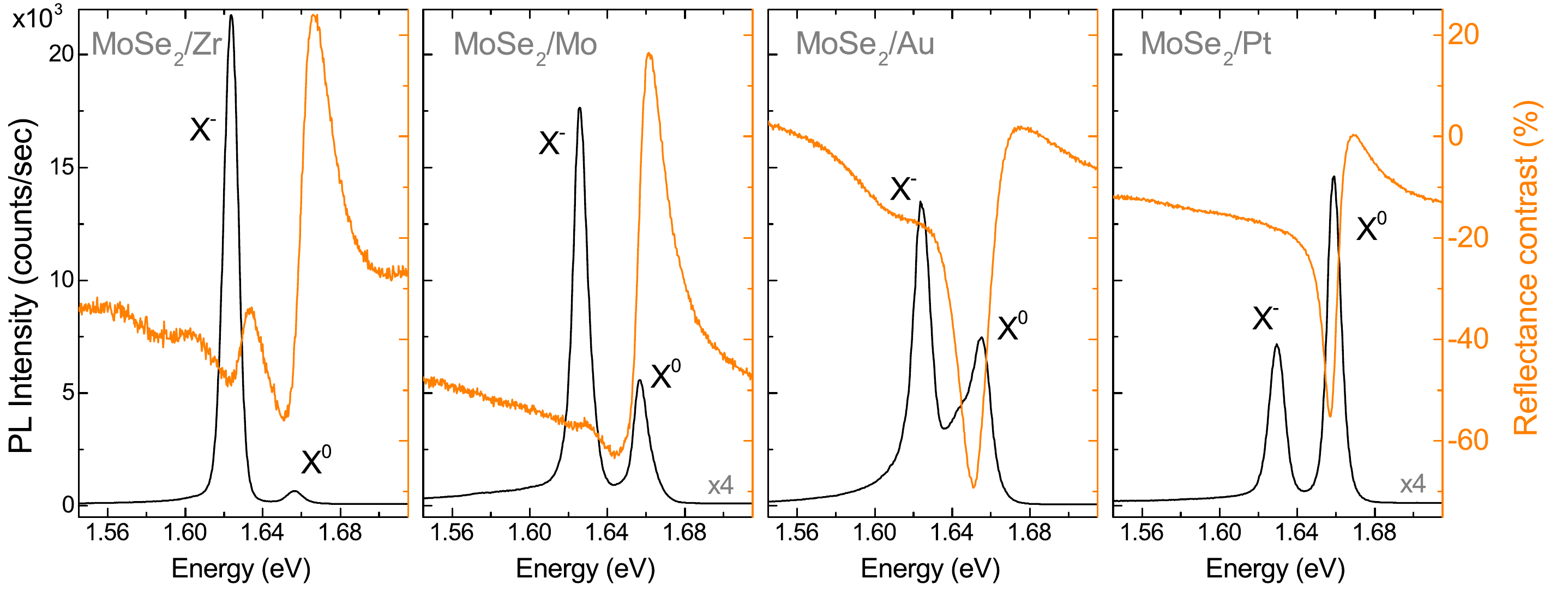}
\caption{Photoluminescence (PL) and reflectance contrast (RC) spectra of ML MoSe$_{2}$ deposited on different metallic substrates, measured at T~=~5~K. Note that the vertical scales of the PL intensity and RC are set the same for all four panels.}
\label{fig:PLRC}
\end{figure*}

In this paper, we study the effect of metallic surrounding on the optical properties of monolayer MoSe$_{2}$. The ground exciton state of the MoSe$_{2}$ ML is bright \cite{kormanyos2015k} and the corresponding emission comprises two peaks related to neutral and charged excitons\cite{arora2015exciton, koperski2017optical}. Metals, on top of which the MoSe$_{2}$ flakes were transferred, were chosen based on their fundamental physical properties: electrical and thermal conductance, work functions and chemical stability. Platinum (Pt) and gold (Au) are often used as high-work-function electrical contacts (the work functions of Pt and Au are equal to 5.64~eV \cite{Pt} and 5.1~eV \cite{metal_WF}, respectively). When connected to monolayer MoSe$_2$ they are expected to form p-type Schottky barriers (for a schematic presentation of the energy structure of 1 ML MoSe$_2$ metal junctions under study see Fig.\ref{fig:probka}(a)). The opposite should be observed for zirconium (Zr), characterised by low work function (equal to 4.05~eV \cite{metal_WF}) and supposed to result in n-type Schottky contacts. We also consider molybdenum (Mo) that should form strong orbital overlaps with materials comprising the same element, in particular, MoSe$_2$.


\section{Samples and experimental setups \label{sec:methods}}

Metallic substrates were prepared by laser lithography and e-beam evaporation employed for patterning pieces of an Si/(90~nm)SiO$_{2}$ wafer with a network of slabs made of 5 nm thick Ti adhesion layer followed by 25~nm thick Pt, Mo, Au, or Zr layer. MoSe$_{2}$ monolayers were prepared by all-dry PDMS-based exfoliation \cite{detertrans} of bulk crystals purchased from HQ Graphene. The flakes of interest were first identified under an optical microscope and then subjected to atomic force microscopy and Raman spectroscopy characterisation to unambiguously determine their thicknesses and assess their overall quality. Right before transferring the flakes onto a chosen substrate, the substrate's surface was cleaned with either dry CHF$_{3}$ reactive-ion-plasma (Pt, Au, Mo) or wet HF etching (Zr) to remove the native oxide layer and other possible contaminants. A schematic representation of the samples is shown in Fig. \ref{fig:probka}(b). To verify the credibility of the obtained results, two sets of samples were produced in the same manner.

The investigated samples were placed on a cold finger of a continuous flow cryostat mounted on x-y motorized positioners. The excitation light was focused through a 50x long-working distance objective with a 0.5 numerical aperture giving a laser spot of about 1~$\upmu$m diameter. The signal was collected via the same microscope objective, sent through a 0.5~m monochromator, and then detected by a CCD camera. The photoluminescence (PL) measurements were carried out using $\lambda$ = 514.5~nm radiation from a continuous wave Ar$^+$ ion laser. The excitation power focused on the sample was kept at $\sim$50~$\upmu$W during all PL measurements to avoid local heating. For the reflectance contrast (RC) study, a 100~W tungsten halogen lamp was used as a source of excitation. Light from the lamp was coupled to a multimode fiber of 50~$\upmu$m core diameter, and then collimated and focused on the sample to a spot of about 4~$\upmu$m diameter. All measurements were performed over a wide range of temperatures from 5~K to 300~K. The unpolarized Raman scattering measurements were carried out in the backscattering geometry at room temperature using a $\lambda$ = 532~nm CW diode laser. The power of light on the samples did not exceed 70~$\upmu$W. The collected Raman signal was dispersed by a 0.75~m spectrometer equipped with 1800 grooves/mm gratings.


\section{Results \label{sec:results}}

\begin{figure*}
\includegraphics[width=\linewidth]{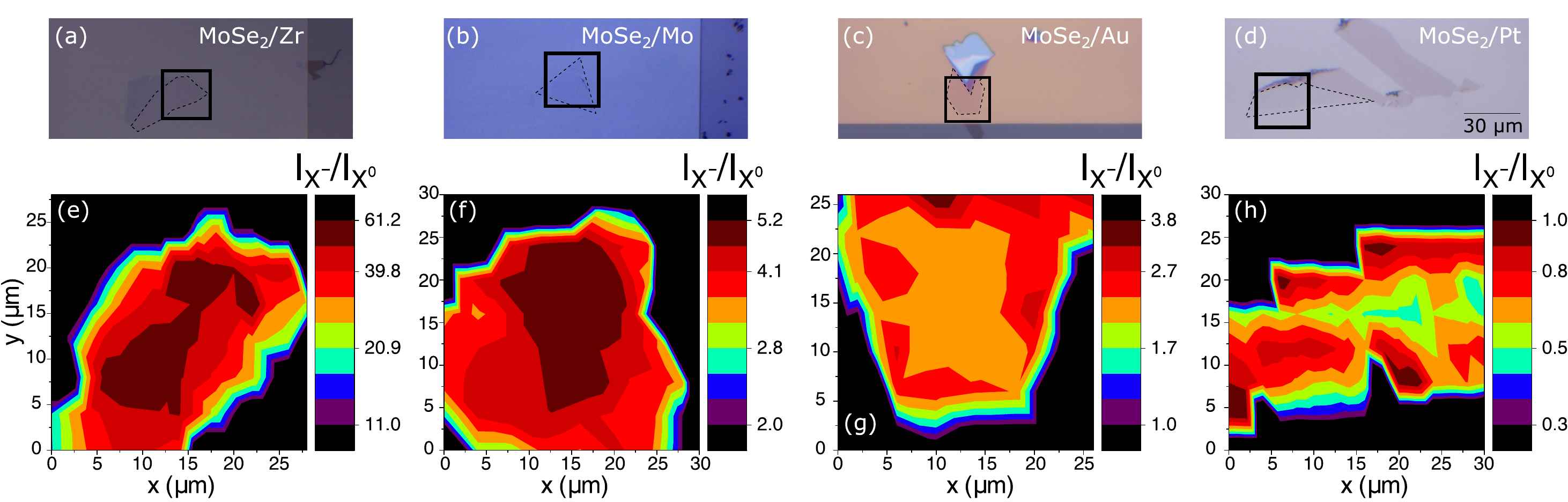}
\caption{(a)-(d) Optical images of investigated flakes. Dashed lines indicate the boundaries of MoSe$_2$ MLs and solid squares represent the mapped area. (e)-(h) False-colour spatial maps of an intensity ratio of the trion line to the neutral exciton line (X$^-$/X$^0$), measured T~=~5~K on MoSe$_{2}$ MLs deposited on different metallic substrates. Note that the color scale is logarithmic.}
\label{fig:maps}
\end{figure*}

Room-temperature (RT) Raman scattering spectra measured on the studied structures are presented in Fig. \ref{fig:probka}(c). The spectra display two modes: an intra-layer $\mathrm{E'}$ at 240~cm$^{-1}$ and an out--of--plane $\mathrm{A'_{1}}$ mode at $\sim$290~cm$^{-1}$. These are characteristic results of MoSe$_{2}$ MLs \cite{tonndorf}, which confirms the single-layer thickness of our samples. The Raman spectra are not significantly affected by either the metal on which the flakes were deposited or strain and disorder that could have been introduced into the flakes during the fabrication process. 

The PL spectra, shown in Fig. \ref{fig:PLRC} with black lines, are composed of two well-separated emission which are attributed to the neutral (X$^0$ $\sim$1.66~eV) and negatively charged (X$^-$ $\sim$1.63~eV) excitons formed in the vicinity of the so-called A exciton at the K$^\pm$ points of the Brillouin zone \cite{arora2015exciton,molas2017brightening}. The identification of the negatively or positively charged excitons is not straightforward. There are, however, strong arguments for n-type doping of the studied MLs, which are presented in the following section. Additionally, a third feature with a maximum at around 1.65~eV can be observed in the PL spectrum collected on the ML deposited on the Au substrate. A similar emission peak has not been reported so far for MoSe$_2$ MLs. A possible assignment of this peak is difficult as the contribution of phonons, dark excitons and biexcitons is not very likely. By comparing the spectra (with panels arranged by increasing metal work function from left to right) an obvious trend can immediately be noticed. With increasing work function of the metal, the relative intensity of the neutral excitonic line to the charged exciton line increases. For the MoSe$_{2}$/Zr structure, the emission related to the neutral exciton X$^{0}$ is about 30 times weaker than that of the charged exciton. On the other hand, the intensities of the X$^-$ lines are about three and two times larger as compared to the X$^0$ peaks for the MoSe$_{2}$/Mo and MoSe$_{2}$/Au structures, respectively. In the case of MoSe$_{2}$/Pt structure, for which the metal work function is the highest, the neutral exciton emission is about two times stronger as compared to the charged exciton one. An analogous effect can be recognized in the RC results measured at T~=~5~K, shown in Fig. \ref{fig:PLRC} with orange lines. For three structures, $i.e.$ MoSe$_{2}$/Zr,  MoSe$_{2}$/Mo and MoSe$_{2}$/Au, two resonances can be observed in the RC spectra, which are attributed to the charged and neutral excitons \cite{arora2015exciton,koperski2017optical,koperski2018orbital}. For the MoSe$_{2}$/Pt stack, there is only one dip in the RC spectrum, which is ascribed to the neutral exciton.

To investigate the homogeneity of our samples, we mapped their PL signal over the areas marked in Fig.~\ref{fig:maps}(a)-(d) with solid black squares. The measurements were done at T=5~K. Shown in the maps is the intensity ratio of the trion line to the neutral exciton line (X$^-$/X$^0$). The analysis of the obtained X$^-$/X$^0$ ratios for different substrates leads to the following conclusions: (i) the ratios are significantly affected by the position on the flake, $i.e.$ they change between the center and the edge of a ML by about 2-6 times, which implies that the doping level in a given ML is smallest at its edge, (ii) for three MLs, $i.e.$ those deposited on Zr, Mo, and Au, quite big regions of homogenous intensity ratios are observed, and (iii) in the case of the ML placed on Pt, the inhomogeneity is much larger (it most probably originates from irregular flake's shape), but still the mapped region contains an about $10 \times 10~\upmu\textrm{m}$ area where the X$^-$/X$^0$ ratio is close to 0.5 or smaller.

\begin{figure*}
\includegraphics[width=\linewidth]{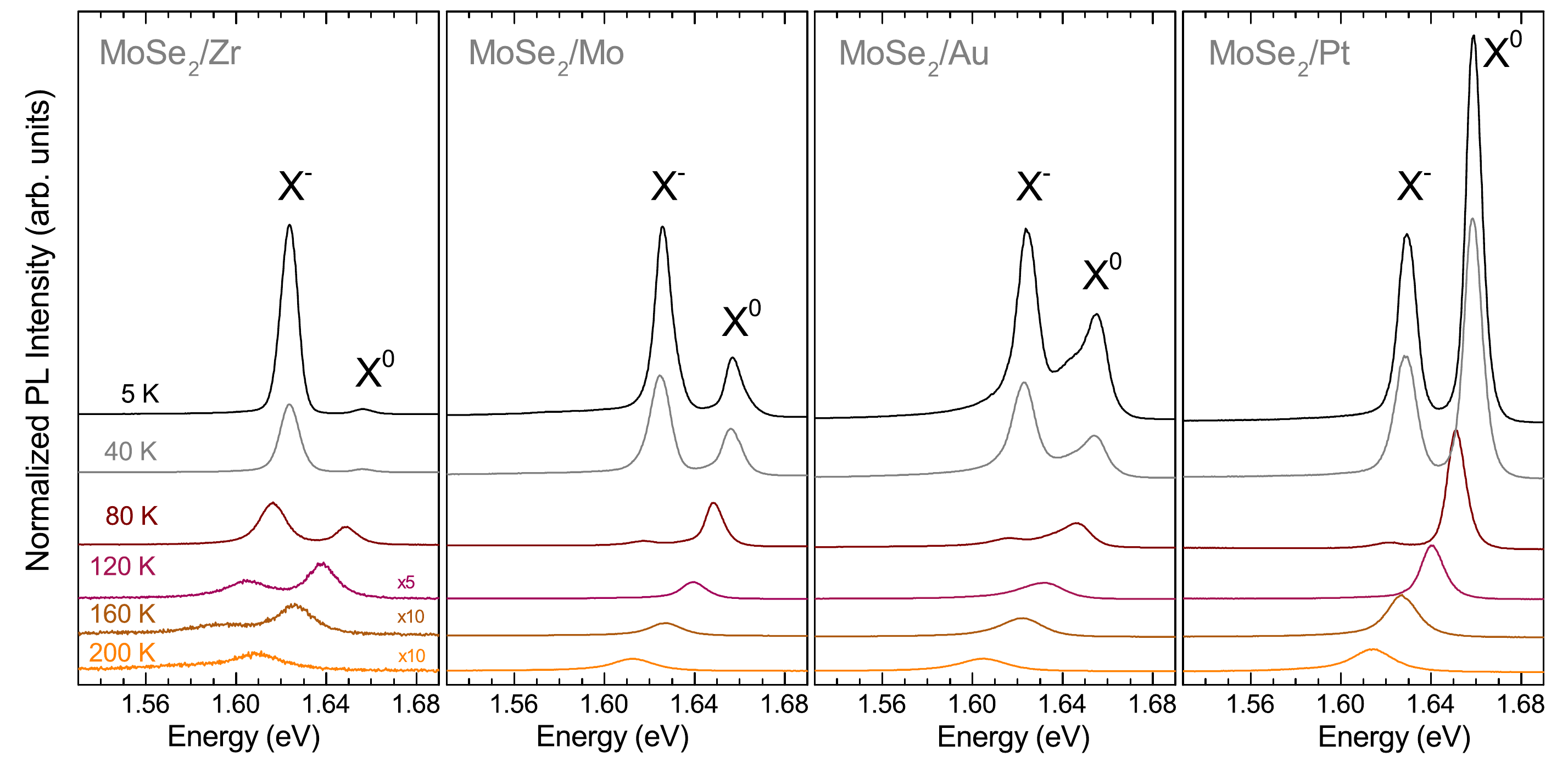}
\caption{Temperature evolution of PL spectra of MoSe$_{2}$ MLs deposited on different metallic substrates. The PL spectra are normalized to the intensity of the X$^-$ line at 5~K. The spectra are vertically shifted for clarity and some of them are multiplied by scaling factors in order to avoid their intersections with the neighbouring experimental curves. }
\label{fig:PLnorm}
\end{figure*}

To examine the emission properties of studied MoSe$_2$ MLs in more detail, we subjected them to PL measurements performed over a wide temperature range from 5~K to 300~K. It is known that increasing the temperature of typical MoSe$_2$ MLs deposited on Si/SiO$_2$ leads to quickly vanishing X$^-$ emission \cite{arora2015exciton}. Selected PL spectra from our temperature study are shown in Fig. \ref{fig:PLnorm}. In order to maintain the legibility of the plot, the curves are displaced vertically and, if needed, multiplied by a scaling factor. Two main effects of the temperature can be distinguished. At low temperature, the PL spectrum of the MoSe$_{2}$/Zr sample is dominated by trion's contribution, which then rapidly quenches as the temperature increases and can not be observed at T$>$200~K. In the case of three other structures, $i.e.$ MoSe$_{2}$/Mo,  MoSe$_{2}$/Au and MoSe$_{2}$/Pt, the X$^-$ emission disappears from the PL spectra more quickly and can not be recognized at T$>$120~K. Finally, for all studied structures, only the X$^{0}$ line is apparent in the PL spectra at T$>$200~K. The X$^{0}$-exciton feature shows an overall redshift consistent with the temperature evolution of the band gap \cite{arora2015exciton}. 

\section{Discussion  \label{sec:Discussion}}

The observed effects of the metallic substrates on the optical response of the MoSe$_2$ ML can be explained in terms of corresponding doping levels. A schematic representation of energy levels of the studied ML and the metals used as substrates is presented in Fig. \ref{fig:probka}(a). As can be seen in the Figure \ref{fig:probka}(a), the relative position of the Fermi levels in metals with respect to the conduction and valence bands edges in the MoSe$_2$ ML changes significantly due to the variation of the metal work function. For the two metals characterised by extreme work functions, $i.e.$ Zr and Pt, their levels coincide correspondingly with the conduction and valence bands of the MoSe$_2$ monolayer. It may result in the creation of metal/semiconductor junctions, which exhibit n- (Zr) or p-type (Pt) characteristics and therefore permit to form, respectively, the negatively or positively charged excitons. This observation is in good agreement with data that has previously been reported for TMD/metal interfaces \cite{FL,pan2016interfacial,2d} . 

\begin{figure}
\includegraphics[width=\linewidth]{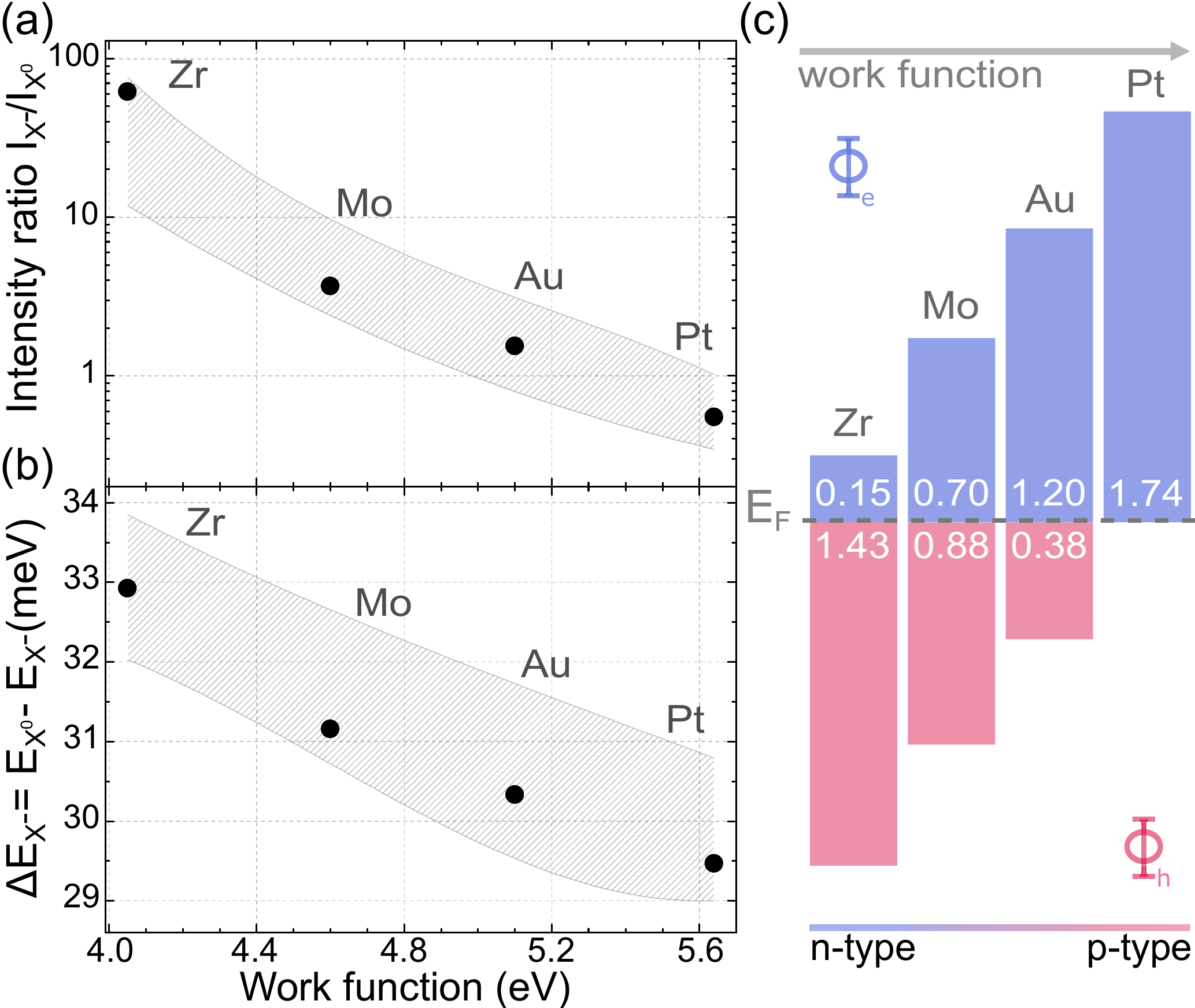}
\caption{(a) Intensity ratio of the charged exciton to the neutral exciton line and (b) charged exciton binding energy ($\Delta{E_{X^{-}}}$) versus the metal work function. The grey-shaded areas represent the deviation range of the values marked with solid circles and extracted from Fig. \ref{fig:PLRC}, obtained by analysing the whole spatial maps shown in Fig.~\ref{fig:maps}. (c) Comparison of calculated Schottky Barrier Heights ($\Phi$) for selected metal/MoSe$_{2}$ junctions.}
\label{fig:wf}
\end{figure}

In order to thoroughly investigate the observed effect of metallic substrate on the charged excitons in the studied MLs, we performed an analysis of the energy positions of the conduction and valence bands of MoSe$_2$ ML as compared to the metals' work functions. Similar energy values of MoSe$_{2}$ affinity and Zr work function result in the band alignment. The electrons can be easily transferred between the MoSe$_{2}$ conduction band and the metal surface, shifting the Fermi level to higher energies. In this case, the structure can be characterised as a Schottky barrier, which serves as an efficient electron trap. As a consequence of this band alignment, we expect that the studied ML deposited on Zr reveals relatively high n-type doping. This leads to the appearance of the negatively charged excitons in both the PL and RC spectra (see Fig. \ref{fig:PLRC}). The high doping level in the MoSe$_{2}$/Zr structure results in the observation of the X$^-$ resonance in the corresponding RC spectrum measured at T=5~K (see Fig.~\ref{fig:PLRC}). The MoSe$_{2}$ MLs on Mo and Au are less n-type doped, but still two X$^{0}$ and X$^-$ resonances can be recognized in both their RC and PL spectra (see Fig. \ref{fig:PLRC}). In these two cases (Mo and Au), the Fermi energy of the metal is located inside the band gap of the MoSe$_2$ monolayer. Assuming that the exfoliated MoSe$_2$ crystals were intentionally undoped, their Fermi level should be in the middle of the energy gap as in conventional semiconductors, which would amount to about 4.69 eV,  $i.e.$ close to the work functions of Mo (4.6~eV) and Au (5.1~eV). Consequently, it is expected that the MoSe$_{2}$ MLs remain essentially undoped when placed on Mo or Au substrates, and only the neutral exciton resonance may be observed in the RC and PL spectra. As can be seen in Fig. \ref{fig:PLRC}, the X$^{0}$ and X$^{-}$ transitions are apparent in both types of experiments, which strongly suggests that the exfoliated MoSe$_2$ crystals were unintentionally n-doped. Note that the measured PL spectra of MoSe$_2$ deposited on Zr, Mo and Au substrates resemble those of typically studied MoSe$_2$ samples on Si/SiO$_2$ substrates \cite{arora2015exciton, wang2015exciton, kioseoglou2016optical, molas2017brightening}. The spectra of the MoSe$_{2}$/Pt structure show that the neutral exciton emission is more intense than the trion one. As platinum's work function falls in the valence band of the investigated ML, the formation of the p-type doping in the MoSe$_2$ ML can be expected in such a case. However, as we already discussed, the MoSe$_2$ crystals used for exfoliation were probably unintentionally n-doped. The deposition of the ML on the Pt substrate results in a significant decrease of the X$^-$ intensity, but do not permit to create positively charged excitons. Moreover, as it was shown in Ref. \citenum{ross2013electrical}, the binding energy of the negative trion is affected by electrostatically-tuned doping level to a larger extent than the binding energy of the positive trion, which may also support our attribution of the lower energy feature in the PL spectra to the negative trion (see Fig. \ref{fig:wf}(b)).

Figures \ref{fig:wf}(a) and (b) present an intensity ratio (I$_{X^{-}}$/I$_{X^{0}}$) and energy difference ($\Delta E_{X^{-}} = E_{X^{0}} - E_{X^{-}}$) of the neutral and charged exciton emission lines. Note that $\Delta E_{X^{-}}$ can be defined as the binding (dissociation) energy of the charged exciton, which is the energy required to promote one of the trion's electrons to the conduction band edge in the limit of infinitesimally small doping \cite{huard2000bound, mak2013tightly}. As can be seen in that Figure \ref{fig:wf}, both the intensity ratio and the trion's binding energy systematically decrease with the increase of the metal work function. However, the quantitative impact of the work functions on the observed changes, shown in Figs. \ref{fig:wf}(a) and (b), varies considerably. While the trion binding energy changes by about 10\% with increasing the work function, the intensity ratio decreases by more than 50 times. It is important to mention that the influence of the metallic substrate on the trion binding energy is probably accompanied by variation of the neutral exciton binding energy ($\Delta E_{X^{0}}$), similarly as it was demonstrated for different dielectric environments of S-TMD monolayers~\cite{raja,molas2019PRL}. However, a recent theoretical work \cite{hichri} demonstrates that the ratio of the trion to the exciton binding energy ($\Delta E_{X^{-}}$/$\Delta E_{X^{0}}$) is not fixed, but changes with the environment of the ML. In consequence, we are not able to quantitatively estimate the effect of metallic substrate on the neutral exciton binding energy. 

In many practical cases, metal-semiconductor junctions can be reasonably described by a simple model relying on the Schottky Barrier Height ($\Phi$), an energy of the charge carriers have to overcome while being transported across the junction. The possibility of tuning $\Phi$ is highly desirable for various reasons, most of which determine the quality of electronic devices based on TMDs, especially from the viewpoint of the reduction of contact resistance \cite{xu2016universal}. The biggest difficulty in constructing efficient electrical contacts to TMD layers is so-called Fermi level pinning (FLP) \cite{das2012high,bampoulis2017defect}. The strength of FLP in a given semiconductor brought into contact with a set of metals of different work functions can be characterised by a slope of linear dependence fitted to the $\Phi$-versus-$\chi$ data. In our case, we neglect the contribution from this effect by assuming weak interactions between the metal and the MoSe$_{2}$ ML \cite{liu2016van,pan2016interfacial,ouyang2018tunable}. By using the Schottky-Mott model it is straightforward to calculate the Schottky Barrier Heights for various metal/semiconductor junctions: 
\begin{align*}
        \Phi_e &= \Phi-\chi \\
    \Phi_h &= E_{ip}-\Phi
\end{align*}
where $\Phi_e$ and $\Phi_h$ are the barrier heights for electrons and holes, respectively, $\chi$ is the semiconductor electron affinity and $E_{ip}$ denotes the ionization potential. The obtained values are presented in Fig. \ref{fig:wf}(c). Our results show good agreement with the above analysis based on the relative alignment of the conduction and valence bands in the MoSe$_2$ MLs and metals' work function sketched in Fig. \ref{fig:probka}. As can be appreciated in Fig. \ref{fig:wf}, the lowest Schottky barrier height for electrons of about 0.15 eV is obtained for Zr, while the highest one, equal to about 1.74 eV, for Pt. Interestingly, the Schottky barrier height for holes is almost 0 for Pt. These simple calculations support our speculations based on experimental results, that the type of doping in MoSe$_2$ monolayer can be altered in a controlled way by placing it on metallic substrates with selected work functions.

\section{Conclusions \label{sec:Conclusions}}

We have investigated the effect of metallic (Pt, Au, Mo, or Zr) substrate on the optical response of monolayer MoSe$_{2}$. It has been found that the emission intensity ratio of the neutral to charged excitons as well as the trion binding energy decrease with increasing the work function of the substrate. Our measurements revealed that the PL and RC spectra of the structure expected to exhibit p-type characteristics (MoSe$_{2}$/Pt) are dominated by the neutral exciton. When the Fermi level of metals falls inside the MoSe$_2$ ML band gap, like for Mo and Au, both the PL and RC spectra show two resonances due to the neutral and charged excitons. On the contrary, in the structure with the metal's work function matching the bottom of the semiconductor's conduction band (MoSe$_{2}$/Zr) strong resonances originating from the negatively charged exciton are seen in both the PL and RC spectra. We explain this effect in terms of variable doping of the MoSe$_{2}$ ML induced by the metal substrate. The alignment of the energy levels at the metal/semiconductor junction allows for the transfer of carriers between the layers. The presented results demonstrate a doping method of monolayer TMDs by appropriately selecting the metallic substrates as well as versatility of standard optical experimental methods like photoluminescence and reflectance contrast, which can be successfully used to check the quality and characteristics of metal/semiconductor junctions.


\begin{acknowledgments}
The work has been supported by the National Science Centre, Poland (grant no. 2017/27/B/ST3/00205, 2017/27/N/ST3/01612, 2018/31/B/ST3/02111), the ATOMOPTO project (TEAM programme of the Foundation for Polish Science co-financed by the EU within the ERDFund), the EU Graphene Flagship project (No. 785219), the Nanofab facility of the Institut N\'eel, CNRS, and Ministry of Education, Youth and Sports of the Czech Republic under the project CEITEC 2020 (LQ1601).
\end{acknowledgments}

\bibliographystyle{apsrev4-1}
\bibliography{bib}

\end{document}